\documentclass[preprint,prb]{revtex4}

\usepackage{amsmath}
\usepackage{amssymb}
\usepackage{amsthm}
\newcommand{{\bfr}}{\mbox{\boldmath$r$\unboldmath}}
\newcommand{{\bfv}}{\mbox{\boldmath$v$\unboldmath}}
\newcommand{{\bff}}{\mbox{\boldmath$f$\unboldmath}}
\newcommand{{\bfF}}{\mbox{\boldmath$F$\unboldmath}}
\newcommand{{\bfA}}{\mbox{\boldmath$A$\unboldmath}}
\newcommand{\gradv}{\boldsymbol{\nabla}}
\def\v#1{{\bf#1}}

\begin{document}

\title{Can Maxwell's equations be obtained from\\ the continuity equation?}

\author{Jos\'e A. Heras}
\email{herasgomez@gmail.com}

\affiliation{Departamento de F\'isica y Matem\'aticas, Universidad Iberoamericana,
Prolongaci\'on Paseo de la Reforma 880, M\'exico D. F. 01210, M\'exico}
\affiliation{
Department of Physics and Astronomy, Louisiana State University, Baton
Rouge, Louisiana 70803-4001, USA}

\begin{abstract}
We formulate an existence theorem that states that given localized scalar and vector time-dependent sources satisfying the continuity equation, there exist two retarded fields that satisfy a set of four field equations. If the theorem is applied to the usual electromagnetic charge and current densities, the retarded fields are identified with the electric and magnetic fields and the associated field equations with Maxwell's equations. This application of the theorem suggests that charge conservation can be considered to be the fundamental assumption underlying Maxwell's equations.
\end{abstract}

\maketitle

\section{INTRODUCTION}
The traditional presentation of Maxwell's equations follows the historical approach in which  electrostatics and magnetostatics are studied first. Then Faraday's induction law is introduced to consider quasistatic phenomena. Finally, we introduce the displacement current to insure charge conservation and obtain the Amp\'ere-Maxwell law, which completes the set of equations known as Maxwell's equations. 

The reader might wonder why an axiomatic approach to Maxwell's equations is not usually presented in undergraduate textbooks, although an axiomatic presentation of quantum mechanics and general relativity, for example, can be found.\cite{1,2} 
One virtue of the axiomatic approach is that it offers the shortest way to the essence of a theory and enables a more rigorous formulation.\cite{3} 
The basic problem for an axiomatic presentation of Maxwell's equations is recognizing the essential postulates underlying Maxwell's equations. One of these postulates is unavoidable: charge conservation, which is expressed by the continuity equation for the charge and current densities. As is well-known, Maxwell's equations imply the continuity equation, but is the converse implication true? If the continuity equation implies Maxwell's equations, then charge conservation should be considered as the fundamental axiom underlying these equations.

In this paper we show how Maxwell's equations can be obtained from the continuity equation. We formulate an existence theorem that 
states that given localized time-dependent scalar and vector sources satisfying the continuity equation, there exist two retarded fields that satisfy a set of four coupled field equations. When the theorem is applied to the usual electromagnetic charge and current densities, the retarded fields are identified with the electric and magnetic fields in the form given by Jefimenko\cite{4} and the associated field equations are naturally identified with Maxwell's equations. Therefore, not only do Maxwell's equations imply the continuity equation, but the continuity equation implies Maxwell's equations. The application of the theorem to electromagnetic sources suggests an axiomatic approach to Maxwell's equations in which charge conservation is considered to be the fundamental axiom underlying these equations.

\section{Existence theorem for two retarded fields}

In this section we formulate and demonstrate the following existence theorem: Given the localized sources $\rho(\v x,t)$ and $\v J(\v x,t)$ which satisfy the continuity equation, 
\begin{equation}
\gradv\cdot\v J+\frac{\partial\rho}{\partial t}=0,
\end{equation}
there exist retarded fields $\v F(\v x,t)$ and $\v G(\v x,t)$ defined by 
\begin{subequations}
\begin{align}
\v F &= \frac{\alpha}{4\pi}\!\int d^3x'\bigg(\frac{\hat{\v R}}{R^2}[\rho]+\frac{\hat{\v R}}{Rc}\left[\frac{\partial \rho}{\partial t}\right]
-\frac{1}{Rc^2}\left[\frac{\partial \v J}{\partial t}\right]\bigg),\\
\v G&= \frac{\beta}{4\pi}\!\int d^3x' \bigg([\v J]\times\frac{\hat{\v R}}{R^2 }+\bigg[\frac{\partial \v J}{\partial t}\bigg]\times\frac{\hat{\v R}}{R c}\bigg).
\end{align}
\end{subequations}
that satisfy the field equations 
\begin{subequations}
\begin{align}
\gradv\cdot\v F&=\alpha\rho,\\ 
\gradv\cdot\v G&= 0,\\
\gradv\times \v F+\gamma\frac{\partial \v G}{\partial t}&= 0,\\
\gradv\times \v G-\frac{\beta}{\alpha}\frac{\partial \v F}{\partial t}
&=\beta\v J.
\end{align}
\end{subequations}
The arbitrary positive constants $\alpha$, $\beta$, $\gamma$, and $c$ are related by $\alpha=\beta\gamma c^2$; 
$\hat{\v R}={\v R}/R =(\v x-\v x')/|\v x-\v x'|$; and the square brackets $[\;]$ indicate that the enclosed quantity is to be evaluated at the retarded time $t'=t-R/c$. 

To prove this theorem, we start by deriving the following identities from Eq.~(1):
\begin{subequations}
\begin{align}
\gradv\times\bigg([\v J]\times\frac{\hat{\v R}}{R^2}+\bigg[\frac{\partial \v J}{\partial t}\bigg]\times\frac{\hat{\v R}}{R c}\bigg)
-\frac{\partial}{\partial t}&\bigg(\frac{\hat{\v R}}{R^2}[\rho]+\frac{\hat{\v R}}{Rc}\left[\frac{\partial \rho}{\partial t}\right]
-\frac{1}{Rc^2}\left[\frac{\partial \v J}{\partial t}\right]\bigg)\nonumber\\
&=4\pi[\v J]\delta(\v x-\v x')-\gradv\left(\gradv'\cdot\frac{[\v J]}{R}\right).\\
\gradv\cdot\bigg(\frac{\hat{\v R}}{R^2}[\rho]+\frac{\hat{\v R}}{Rc}\left[\frac{\partial \rho}{\partial t}\right]
-\frac{1}{Rc^2}\left[\frac{\partial \v J}{\partial t}\right]\bigg)&=4\pi[\rho]\delta(\v x-\v x')+\gradv'\cdot\left(\frac{1}{R c^2}\left[\frac{\partial \v J}{\partial t}\right]\right),
\end{align}
\end{subequations}
where $c$ is arbitrary positive constant, $\delta$ is the Dirac delta function, and $\rho$ and $\v J$ are arbitrary functions of space and time satisfying Eq.~(1) at all points and at all times.

Because Eq.~(1) is satisfied at all points and at all times, we can evaluate it at the source point and the retarded time\cite{5}
\begin{equation}
[\gradv'\cdot\v J]+\left[\frac{\partial \rho}{\partial t}\right]=0.
\end{equation}
A heuristic interpretation of Eq.~(5) is as follows. Consider an observer at a particular location in space who has a watch that reads a particular time. The observer is surrounded by nested spheres, on each of which there is a well-defined retarded time (with respect to the observer). Equation (5) states that the continuity equation holds (or rather, held) on each of those spheres, at the relevant retarded time.\cite{6}

We multiply Eq.~(5) by $1/R$ and use $\partial[\rho]/\partial t=[\partial \rho/\partial t]$ (see the Appendix)
to obtain 
\begin{equation}
\frac{[\gradv'\cdot\v J]}{R}+\frac{\partial}{\partial t}\left(\frac{[\rho]}{R}\right)=0.
\end{equation}
We now substitute the identity (see the Appendix) 
\begin{equation}
\frac{[\gradv'\cdot\v J]}{R}=\gradv\cdot\left(\frac{[\v J]}{R}\right)+\gradv'\cdot\left(\frac{[\v J]}{R}\right),
\end{equation}
into Eq.~(6) and write
\begin{equation}
\gradv\cdot\left(\frac{[\v J]}{R}\right)+\frac{\partial}{\partial t}\left(\frac{[\rho]}{R}\right)=-\gradv'\cdot\left(\frac{[\v J]}{R}\right).
\end{equation}
The gradient of Eq.~(8) is
\begin{equation}
\gradv\left(\gradv\cdot\left(\frac{[\v J]}{R}\right)\right)+\frac{\partial}{\partial t}\gradv\left(\frac{[\rho]}{R}\right)=-\gradv\left(\gradv'\cdot\left(\frac{[\v J]}{R}\right)\right).
\end{equation}
We now consider the identity (see the Appendix):
\begin{equation}
\gradv\left(\gradv\cdot\left(\frac{[\v J]}{R}\right)\right)=\gradv\times\left(\gradv\times\left(\frac{[\v J]}{R}\right)\right) + \frac{1}{Rc^2}\frac{\partial^2[\v J]}{\partial t^2} -4\pi[\v J]\delta(\v x-\v x').
\end{equation}
From Eqs.~(9) and (10) and the property $\partial[\v J]/\partial t=[\partial \v J/\partial t]$, we obtain
\begin{equation}
\gradv\times \left(\gradv\times \left(\frac{[\v J]}{R}\right)\right)+\frac{\partial}{\partial t}\left(\gradv\left(\frac{[\rho]}{R}\right)+\frac{1}{Rc^2}\left[\frac{\partial\v J}{\partial t}\right]\right)=4\pi[\v J]\delta(\v x-\v x')-\gradv\left(\gradv'\cdot\frac{[\v J]}{R}\right).
\end{equation}
With the aid of the identities (see the Appendix):
\begin{align}
\gradv\times \left(\frac{[\v J]}{R}\right)&=[\v J]\times\frac{\hat{\v R}}{R^2}+\left[\frac{\partial \v J}{\partial t}\right]\times\frac{\hat{\v R}}{Rc},\\
\gradv\left(\frac{[\rho]}{R}\right)=&-\frac{\hat{\v R}}{R^2}[\rho]-\frac{\hat{\v R}}{Rc}\bigg[\frac{\partial \rho}{\partial t}\bigg],
\end{align}
we can see that Eq.~(11) becomes Eq.~(4a). 
To derive the identity (4b) we take the time derivative of Eq.~(8) and use the property $\partial[\v J]/\partial t=[\partial \v J/\partial t]$ to obtain the expression
\begin{eqnarray}
\gradv\cdot\left(\frac{1}{R}\left[\frac{\partial \v J}{\partial t}\right]\right) + 
\frac{1}{R}\frac{\partial^2}{\partial t^2}[\rho] =-\gradv'\cdot\left(\frac{1}{R}\left[\frac{\partial \v J}{\partial t}\right]\right).
\end{eqnarray} 
We next substitute the identity (see the Appendix):
\begin{equation}
\frac{1}{R}\frac{\partial^2}{\partial t^2}[\rho]=c^2\nabla^2\left(\frac{[\rho]}{R}\right)+4\pi c^2[\rho]\delta(\v x-\v x').
\end{equation}
into Eq.~(14) and obtain after rearranging terms
\begin{equation}
\gradv\cdot\bigg(-\gradv\bigg(\frac{[\rho]}{R}\bigg)-\frac{1}{Rc^2}\left[\frac{\partial \v J}{\partial t}\right]\bigg)= 4\pi[\rho]\delta(\v x-\v x')+\gradv'\cdot\left(\frac{1}{Rc^2}\left[\frac{\partial \v J}{\partial t}\right]\right).
\end{equation}
We then substitute Eq.~(13) into Eq.~(16) to derive Eq.~(4b). 

We next use Eq.~(4) to obtain Eqs.~(2) and (3). Note that the last term on the right-hand side in Eq.~(4), after integrated over all space, can be transformed into a surface integral that vanishes at infinity if the sources are localized, that is, when the sources are zero outside the surface of a finite region of space. Henceforth we will assume that $\rho$ and $\v J$ are localized sources.\cite{7} We multiply the second term on the left-hand side of Eq.~(4a) by $\beta\alpha/(4\pi\alpha)$ and the 
remaining terms by $\beta/(4\pi)$ and integrate over all space: 
\begin{align}
\frac{\beta}{4\pi}\!\int d^3x' \gradv\times\bigg([\v J]\times\frac{\hat{\v R}}{R^2}+&\bigg[\frac{\partial \v J}{\partial t}\bigg]\times \frac{\hat{\v R}}{R c}\bigg)
-\frac{\beta\alpha}{4\pi\alpha}\!\int d^3x' \frac{\partial}{\partial t}\bigg(\frac{\hat{\v R}}{R^2}[\rho]+\frac{\hat{\v R}}{Rc}\bigg[\frac{\partial \rho}{\partial t}\bigg]
-\frac{1}{Rc^2}\bigg[\frac{\partial \v J}{\partial t}\bigg]\bigg)\nonumber\\
&=\beta\!\int d^3x'[\v J]\delta(\v x-\v x')-\frac{\beta}{4\pi}\!\int d^3x'\gradv\bigg(\gradv'\cdot\frac{[\v J]}{Rc}\bigg).
\end{align}
The operators $\gradv\times$ and $\partial/\partial t$ can be extracted outside the integrals of the left-hand side. 
The first term on the right-hand side of Eq.~(17) becomes $4\pi\beta\v J$ after integration over the delta function. 
The operator $\gradv$ can be extracted outside the last integral on the right-hand side, and the resulting integral can be transformed into a surface integral that vanishes at infinity because $\v J$ is localized. Therefore, Eq.~(17) reduces to 
\begin{align}
\gradv\times\bigg\{\frac{\beta}{4\pi}\!\int d^3x'& \bigg([\v J]\times\frac{\hat{\v R}}{R^2}+\bigg[\frac{\partial \v J}{\partial t}\bigg]\times\frac{\hat{\v R}}{R c}\bigg)\bigg\}\nonumber\\
&-\frac{\beta}{\alpha}\frac{\partial}{\partial t}\bigg\{\frac{\alpha}{4\pi}\!\int d^3x'\bigg(\frac{\hat{\v R}}{R^2}[\rho]+\frac{\hat{\v R}}{Rc}\bigg[\frac{\partial \rho}{\partial t}\bigg]
-\frac{1}{Rc^2}\bigg[\frac{\partial \v J}{\partial t}\bigg]\bigg)\bigg\}
=\beta\v J.
\end{align}
We next multiply Eq.~(4b) by $\alpha/(4\pi)$ and integrate over all space to obtain
\begin{align}
\frac{\alpha}{4\pi}\!\int d^3x'\gradv\cdot\bigg(\frac{\hat{\v R}}{R^2}[\rho]+\frac{\hat{\v R}}{Rc}\left[\frac{\partial \rho}{\partial t}\right]
-\frac{1}{Rc^2}\left[\frac{\partial \v J}{\partial t}\right]\bigg) &= \alpha\!\int d^3x'[\rho]\delta(\v x-\v x')\nonumber\\
&+\frac{\alpha}{4\pi}\!\int d^3x'\gradv'\cdot\left(\frac{1}{R c^2}\left[\frac{\partial \v J}{\partial t}\right]\right).
\end{align}
If we follow an argument similar to that used to go from Eq.~(4a) to Eq.~(18), we obtain
\begin{equation}
\gradv\cdot\bigg\{\frac{\alpha}{4\pi}\!\int d^3x'\bigg(\frac{\hat{\v R}}{R^2}[\rho]+\frac{\hat{\v R}}{Rc}\left[\frac{\partial \rho}{\partial t}\right]
-\frac{1}{Rc^2}\left[\frac{\partial \v J}{\partial t}\right]\bigg)\bigg\}=\alpha\rho.
\end{equation}

Equations (18) and (20) are the main result of this paper. These equations are actually Eqs.~(3d) and (3a) because the quantities within the brackets $\{~\}$ in Eqs.~(18) and (20) are the fields $\v F$ and $\v G$ defined by Eq.~(2). In other words, Eqs.~(18) and (20) 
show the existence of the fields $\v F$ and $\v G$ in terms of which these equations can be written as Eqs.~(3d) and (3a).

To complete the  demonstration of the theorem we need to derive Eqs.~(3b) and (3c). 
We first take the divergence of Eq.~(2b)
\begin{equation}
\gradv\cdot \v G=\frac{\beta}{4\pi}\!\int d^3x' \gradv\cdot\left([\v J]\times\frac{\hat{\v R}}{R^2 }\right)
+\frac{\beta}{4\pi}\!\int d^3x' \gradv\cdot\left(\left[\frac{\partial \v J}{\partial t}\right]\times\frac{\hat{\v R}}{R c}\right).\\
\end{equation}
A direct calculation gives
\begin{subequations}
\begin{align}
\gradv\cdot\left([\v J]\times\frac{\hat{\v R}}{R^2}\right)&= \frac{\hat{\v R}}{R^2}\cdot\gradv\times [\v J]- \left[\v J\right]\cdot\gradv\times\left(\frac{\hat{\v R}}{R^2 c}\right)\\
&= \frac{\hat{\v R}}{R^2}\cdot\left(\left[\frac{\partial \v J}{\partial t}\right]\times\frac{\hat{\v R}}{c}
\right)-\left[\v J\right]\cdot\gradv\times\left(\frac{\hat{\v R}}{R^2 c}\right) \\
&= 0,
\end{align}
\end{subequations}
where Eq.~(A10) and the results $\hat{\v R}\cdot(\v A\times\hat{\v R})= 0$ and $\gradv\times(f(R)\hat{\v R}) = 0$ have been used.
By a similar calculation we can show 
\begin{equation}
\gradv\cdot\left(\left[\frac{\partial \v J}{\partial t}\right]\times\frac{\hat{\v R}}{R c^2}\right)=0.
\end{equation}
If we substitute Eqs.~(22) and (23) into Eq.~(21), we obtain Eq.~(3b). 

We now take the curl of Eq.~(2a)
\begin{equation}
\gradv\times\v F =\frac{\alpha}{4\pi}\!\int d^3x'\gradv\times \left(\frac{\hat{\v R}}{R^2}[\rho]+\frac{\hat{\v R}}{Rc}\left[\frac{\partial \rho}{\partial t}\right]\right)- \frac{\alpha}{4\pi}\!\int d^3x'\gradv\times\left(\frac{1}{Rc^2}\left[\frac{\partial \v J}{\partial t}\right]\right).
\end{equation}
A direct calculation gives
\begin{subequations}
\begin{align}
\gradv\times \left(\frac{\hat{\v R}}{R^2}[\rho]\right)&= -\frac{\hat{\v R}}{R^2}\times\gradv [\rho]+ [\rho]\gradv\times\left(\frac{\hat{\v R}}{R^2 }\right) \\
&= \frac{\hat{\v R}}{R^2}\times\frac{\hat{\v R}}{c}\bigg[\frac{\partial \rho}{\partial t}\bigg]+ [\rho]\gradv\times\left(\frac{\hat{\v R}}{R^2 }\right) =0,
\end{align}
\end{subequations}
where Eq.~(A12) and the results $\hat{\v R}\times\hat{\v R} = 0$ and $\gradv\times(f(R)\hat{\v R}) = 0$ have been considered. 
By a similar calculation we can show 
\begin{equation}
\gradv\times \left(\frac{\hat{\v R}}{Rc}\left[\frac{\partial \rho}{\partial t}\right]\right)=0.
\end{equation}
We now calculate
\begin{subequations}
\begin{align}
\gradv\times\left(\frac{1}{Rc^2}\left[\frac{\partial \v J}{\partial t}\right]\right)&= \frac {1}{c^2}\frac{\partial}{\partial t}
\gradv\times \frac{[\v J]}{R} \\
&= \frac {1}{c^2}\frac{\partial}{\partial t}\left([\v J]\times\frac{\hat{\v R}}{R^2}+\bigg[\frac{\partial \v J}{\partial t}\bigg]\times\frac{\hat{\v R}}{Rc}\right) \\
&= \frac {\beta}{c^2\beta}\frac{\partial}{\partial t}\left([\v J]\times\frac{\hat{\v R}}{R^2}+\bigg[\frac{\partial \v J}{\partial t}\bigg]\times\frac{\hat{\v R}}{Rc}\right).
\end{align}
\end{subequations}
where Eq.~(11) has been used. From Eqs.~(24)--(27) and Eq.~(2) we obtain Eq.~(3c). This result completes the demonstration of the theorem. Note that another theorem that leads to Maxwell's equations has also been recently proposed.\cite{8,9} 

\section{Maxwell's equations and JEFIMENKO'S EQUATIONS}

If we identify $\rho$ and $\v J$ with the usual charge and current densities of electromagnetism and $c$ with the speed of light in vacuum, then Eq.~(2) become Jefimenko's equations\cite{4}
\begin{subequations}
\begin{align}
\v E &= \frac{\alpha}{4\pi}\!\int d^3x'\bigg(\frac{\hat{\v R}}{R^2}[\rho]+\frac{\hat{\v R}}{Rc}\left[\frac{\partial \rho}{\partial t}\right]
-\frac{1}{Rc^2}\left[\frac{\partial \v J}{\partial t}\right]\bigg),\\
\v B&= \frac{\beta}{4\pi}\!\int d^3x' \bigg([\v J]\times\frac{\hat{\v R}}{R^2 }+\bigg[\frac{\partial \v J}{\partial t}\bigg]\times\frac{\hat{\v R}}{R c}\bigg),
\end{align}
\end{subequations}
and Eq.~(3) becomes Maxwell's equations,
\begin{subequations}
\begin{align}
\gradv\cdot\v E&=\alpha\rho,\\ 
\gradv\cdot\v B&= 0,\\
\gradv\times \v E+\gamma\frac{\partial \v B}{\partial t}&= 0,\\
\gradv\times \v B-\frac{\beta}{\alpha}\frac{\partial \v E}{\partial t}
&=\beta\v J.
\end{align}
\end{subequations}

Equations (28) and (29) are expressed in a generalized system of units defined by $\alpha$, $\beta$, and $\gamma$ satisfying the relation 
\begin{equation}
\alpha=\beta\gamma c^2. 
\end{equation}
This generalized system contains three systems of units: Gaussian, SI, and Heaviside-Lorentz. The specific values of $\alpha$, $\beta$ and $\gamma$ are given in Table~\ref{tab1}.
If we adopt SI units, then Eqs.~(28) and (29) become Jefimenko's equations\cite{4} and Maxwell's equations are expressed in their usual form.\cite{4} 

\begin{table}[h]
\begin{center}
\begin{tabular}{|l|l|l|l|}
\hline
System & $\;\;\alpha$ & $\;\;\beta$ & $\;\;\gamma$\\
\hline
Gaussian & $\;4\pi$ & $\;4\pi/c $ & $\;1/c$\\ \hline
SI & $\;\;1/\epsilon_0$&$\;\;\mu_0$&\; 1\\ \hline
Heaviside-Lorentz & $\;\;\;1$ & $\;\;1/c$ & $\;\;1/c$\\ 
\hline
\end{tabular}
\end{center}
\caption{\label{tab1}Generalized system of electromagnetic units defined by $\alpha$, $\beta$, and $\gamma$ with $\alpha=\beta\gamma c^2$. This system contains the Gaussian, SI, and Heaviside-Lorentz units.}
\end{table}

\section{Discussion}
The existence theorem for the retarded fields $\v F$ and $\v G$ in Eq.~(2) is a purely mathematical result, that is, the theorem only ascribes mathematical existence to these fields by showing that they satisfy the field equations (3). To ascribe a physical meaning to these abstract fields, we must identify the conserved sources as the electric charge and current densities, and associate the fields 
with the electric and magnetic fields $\v E$ and $\v B$, via the Lorentz force law, $\v F=
e(\v E+\v v/c\times \v B)$. In other words, real electric and magnetic fields produce the observed forces on electric charges which are described by the Lorentz force. Therefore, the physical existence of the fields $\v E$ and $\v B$ can be verified by testing the Lorentz force, which itself constitutes an independent postulate.

The theorem formulated here is formally correct. However, a referee pointed out that the physical implication of the theorem is not surprising and 
posed the following interesting question: what would it mean if the theorem were false from a physical point of view? The referee speculated that given the sources $\rho$ and $\v J$ 
satisfying the continuity equation, they would not yield the retarded fields $\v E$ and $\v B$. We can imagine two situations. 
In one case we can conceive abstract charge and current densities $\widetilde{\rho}$ and $\widetilde{\v J}$ satisfying the continuity equation $\gradv\cdot\widetilde{\v J}+\partial\widetilde{\rho}/\partial t=0$ which produce the instantaneous electric and magnetic fields (in Gaussian units)
\begin{subequations}
\begin{align}
\widetilde{\v E}(\v x,t) &=\!\int d^3x'\frac{\hat{\v R}}{R^2} \widetilde{\rho}(\v x',t),\\
\widetilde{ \v B}(\v x,t)&=\!\int d^3x'\frac{ \widetilde{\v J}(\v x',t)\times\hat{\v R}}{R^2c},
\end{align}
\end{subequations}

\noindent that satisfy the field equations of a Galilean electromagnetic theory\cite{12}
\begin{subequations}
\begin{align}
\gradv\cdot \widetilde{\v E}&=4\pi\widetilde{\rho},\\
\gradv\cdot \widetilde{\v B}&=0,\\
\gradv\times \widetilde{\v E}&=0,\\
\gradv\times \widetilde {\v B}-\frac{1}{c}\frac{\partial \widetilde {\v E}}{\partial t}
&=\frac{4\pi}{c} \widetilde{\v J}.
\end{align}
\end{subequations}
Our problem is that we do not know how to prepare the sources $\widetilde{\rho}$ and $\widetilde{\v J}$ in the 
laboratory. For example, we do not know how to produce sources $\widetilde{\rho}$ and $\widetilde{\v J}$ associated with a moving electron that would yield instantaneous fields. A moving (real) electron always produces retarded electric and magnetic fields satisfying Maxwell's equations, but not instantaneous electric and magnetic fields satisfying the Galilean equations (32). 
We can also imagine that there exist physically realizable charge and current densities satisfying the continuity equation which do not generate electric and magnetic fields (at least, not fields satisfying Maxwell's equations). Because these situations are not possible, we conclude that  given the conserved sources $\rho$ and $\v J$, the fact that we can always construct electric and magnetic fields is not a surprise.

\section{Concluding remarks}
The importance of charge conservation was noted by Maxwell who used it in his discovery of the displacement current. This current allowed him to extend Ampere's law (formulated for a steady-state regime) to the non-stationary regime. Since then, most textbooks introduce the displacement current by invoking charge conservation. But if charge conservation is used to find the final form of Maxwell's equations, then the alert reader might find it surprising that this conservation law can also be considered as a consequence of Maxwell's equations ---the standard proof being that Maxwell's equations directly imply the continuity equation. The idea that charge conservation is not an independent assumption, but a consequence of the laws of electrodynamics is recurrent in textbooks.\cite{10} The reader can see here a circular argument: the continuity equation can be obtained from Maxwell's equations but in obtaining Maxwell's equations, the continuity equation is usually assumed. Our point of view is that charge conservation can be considered as a postulate rather than a result of the theory. Our formal approach to Maxwell's equations based on the existence theorem shows an example of how a single postulate (continuity equation) of a field theory may be used to derive not only the retarded fields associated with that theory, but also the explicit form of the field equations satisfying such fields.

\begin {acknowledgments}
The author is grateful to anonymous referees for their valuable comments and to Professor R.\ F.\ O'Connell for the kind hospitality extended to him in the Department of Physics and Astronomy of the Louisiana State University.
\end{acknowledgments}

\appendix*

\section{identities involving retarded quantities}

We present a simple and non-rigorous derivation of some identities for retarded functions. A retarded function is denoted by $[\rho]$ and is defined as $[\rho]\equiv\rho(\v x',t-R/c)$, where $c$ is an arbitrary positive constant, or equivalently, by $[\rho]\equiv\rho(\v x',t')$, where $t'=t-R/c$. We note that $[\rho]$ depends on the source coordinates not only explicitly, but also implicitly through $R$.
Therefore, the derivation of identities involving space derivatives of retarded quantities is a somewhat complicated task.

We can represent a retarded quantity using the Dirac delta function:
\begin{equation}
[\rho]=\!\int_{-\infty}^\infty\!dt'\,\delta(u)\rho(\v x',t'),
\end{equation}
where  $u=t'-t+R/c$. This representation of a retarded quantity applies also to vector functions: $[\v J]=\!\int\!dt'\,\delta(u)\v J(\v x',t')$. Similar representations can be written for other functions, for example,
$[\gradv'\rho]=\!\int\!dt'\,\delta(u)\gradv'\rho(\v x',t')$; $[\partial \rho/\partial t]=\!\int\!dt'\,\delta(u)\partial \rho(\v x',t')/\partial t'$ and $[\gradv'\cdot \v J]=\!\int\!dt'\,\delta(u)\gradv'\cdot\v J(\v x',t')$. We will now obtain Eqs.~(7), (10), (12), (13), and (15). We will require the following derivatives involving the delta function:
\begin{align}
\frac{\partial \delta (u)}{\partial t}&= -\frac{\partial \delta (u)}{\partial t'},\\
\gradv\delta (u)&= -\gradv'\delta (u),\\
\gradv\delta (u)&= -\frac{\hat{\v R}}{c}\frac{\partial \delta (u)}{\partial t},\\
\nabla^2\delta (u)&= \frac{1}{c^2}\frac{\partial^2 \delta (u)}{\partial t^2} -\frac{2}{Rc}\frac{\partial \delta (u)}{\partial t}. 
\end{align}

We begin by taking the derivative of $[\rho]$ with respect to the present time:
\begin{subequations}
\begin{align}
\frac{\partial [\rho]}{\partial t}&=\!\int dt'\frac{\partial \delta(u)}{\partial t}\rho(\v x',t')= -\!\int dt'\frac{\partial \delta(u)}{\partial t'}\rho(\v x',t' \\
&=\!\int dt'\delta(u)\frac{\partial\rho(\v x',t')}{\partial t'}-\!\int dt'\frac{\partial}{\partial t'}(\delta(u)\rho(\v x',t')) \\
&=\left[\frac{\partial \rho}{\partial t}\right]-\delta(u)\rho(\v x',t')\bigg|_{t'=-\infty}^{t'=-\infty}\nonumber\\
&=\left[\frac{\partial \rho}{\partial t}\right],
\end{align}
\end{subequations}
where we have used Eq.~(A2) and done an integration by parts using the result $\delta(\pm\infty)=0$. By a similar calculation we can show $\partial[\v J]/\partial t=[\partial \v J/\partial t]$.

The divergence of $[\v J]$ with respect to the source coordinates gives
\begin{subequations}
\begin{align}
\gradv'\cdot[\v J] &=\!\int dt'\gradv'\delta(u)\cdot\v J(\v x',t')+\!\int dt'\delta(u)\gradv'\cdot\v J(\v x',t') \\
&= -\!\int dt'\gradv\delta(u)\cdot\v J(\v x',t')+\!\int dt'\delta(u)\gradv'\cdot\v J(\v x',t') \\
&= -\gradv\cdot[\v J]+\left[\gradv'\cdot\v J\right],
\end{align}
\end{subequations}
where we have used Eq.~(A3). By combining the expansions:
\begin{align}
\gradv\cdot\frac{[\v J]}{R} &= \frac{1}{R}\gradv\cdot[\v J]+[\v J]\cdot\gradv\left(\frac{1}{R}\right),\\
\gradv'\cdot\frac{[\v J]}{R} &=\frac{1}{R}\gradv'\cdot[\v J]+[\v J]\cdot\gradv' \left(\frac{1}{R}\right),
\end{align}
and using Eq.~(A7) we obtain Eq.~(7). 

The curl of $[\v J]$ with respect to the field coordinates is
\begin{subequations}
\begin{align}
\gradv\times [\v J] &=\!\int dt'\gradv\delta(u)\times \v J(\v x',t')\\
&=-\frac{\hat{\v R}}{c}\times\frac{\partial}{\partial t}\!\int dt'\delta(u)\v J(\v x',t') \\
&=-\frac{\hat{\v R}}{c}\times\frac{\partial [\v J]}{\partial t}
=\left[\frac{\partial \v J}{\partial t}\right]\times\frac{\hat{\v R}}{c}, 
\end{align}
\end{subequations}
where Eq.~(A4) and $\partial[\v J]/\partial t=[\partial \v J/\partial t]$ have been used. We use Eq.~(A10) and obtain Eq.~(12)
\begin{subequations}
\begin{align}
\gradv\times\frac{ [\v J]}{R} &=\frac{1}{R}\gradv \times[\v J]-[\v J]\times\gradv\left(\frac{1}{R}\right) \\
&= [\v J]\times \frac{\hat{\v R}}{R^2}+\left[\frac{\partial \v J}{\partial t}\right]\times \frac{\hat{\v R}}{Rc}.
\end{align}
\end{subequations}
The gradient of $[\rho]$ with respect to the field coordinates is
\begin{subequations}
\begin{align}
\gradv [\rho] &=\!\int dt'\gradv\delta(u)\rho(\v x',t')\\
&= -\frac{\hat{\v R}}{c}\frac{\partial}{\partial t}\int dt'\delta(u)\rho(\v x',t')\\
&= -\frac{\hat{\v R}}{c}\left[\frac{\partial \rho}{\partial t}\right],
\end{align}
\end{subequations}
where Eqs.~(A4) and (A6) have been used. From Eq.~(A12) it follows that\begin{equation}
\gradv\frac{ [\rho]}{R} = \frac{1}{R}\gradv[\rho]+[\rho]\gradv\left(\frac{1}{R}\right)
= -\frac{\hat{\v R}}{R^2}[\rho]-\frac{\hat{\v R}}{Rc}\left[\frac{\partial \rho}{\partial t}\right].
\end{equation}
The Laplacian of $[\v J]$ with respect to the field coordinates is
\begin{subequations}
\begin{align}
\nabla^2[\v J] &=\!\int dt'\nabla^2\delta(u)\v J(\v x',t') \\
&=\frac{1}{c^2}\frac{\partial^2}{\partial t^2}\int dt'\delta(u)\v J(\v x',t')-\frac{2}{Rc}\frac{\partial}{\partial t}\!\int dt'\delta(u)\v J(\v x',t') \\
&=\frac{1}{c^2}\frac{\partial^2[\v J]}{\partial t^2}-\frac{2}{Rc}\frac{\partial[\v J]}{\partial t},
\end{align}
\end{subequations}
where Eq.~(A5) has been used. Evidently, Eq.~(A14) is valid for the components of $\v J$. For example, if $J_x$ is the component $x$ of $\v J$, then 
\begin{equation}
\nabla^2[J_x]=\frac{1}{c^2}\frac{\partial^2[J_x]}{\partial t^2}-\frac{2}{Rc}\frac{\partial[J_x]}{\partial t}.
\end{equation}
Consider now the Laplacian of $[J_x]/R$ with respect to the field coordinates:
\begin{subequations}
\begin{align}
\nabla^2\frac{[J_x]}{R} &=\frac{1}{R}\nabla^2[J_x]+2\gradv[J_x]\cdot\gradv\left(\frac{1}{R}\right)+[J_x]\nabla^2\left(\frac{1}{R}\right)\\
&=\frac{1}{Rc^2}\frac{\partial^2[J_x]}{\partial t^2}-\frac{2}{R^2c}\frac{\partial[J_x]}{\partial t}+ 2\left(-\frac{\hat{\v R}}{c}\frac{\partial [J_x]}{\partial t} \right)\cdot\left(-\frac{\hat{\v R}}{R^2}\right)- 4\pi[J_x]\delta(\v x-\v x') \\
&=\frac{1}{Rc^2}\frac{\partial^2[J_x]}{\partial t^2}- 4\pi[J_x]\delta(\v x-\v x'),
\end{align}
\end{subequations}
where Eq.~(A12) [with $J_x$ instead of $\rho$] and Eq.~(A15) have been used. Similar expressions for the components $J_y/R$ and $J_z/R$ can be found, and therefore we can write the identity\cite{11}
\begin{equation}
\nabla^2\left(\frac{[\v J]}{R}\right) -\frac{1}{Rc^2}\frac{\partial^2[\v J]}{\partial t^2}= -4\pi[\v J]\delta(\v x-\v x'),
\end{equation}
which yields Eq.~(10) after using the expansion $\nabla^2(~)=\gradv(\gradv\cdot ~)-\gradv\times (\gradv\times ~)$.
Finally, Eq.~(A16) [with $\rho$ instead of $J_x$] directly implies Eq.~(15).


\begin{thebibliography}{99}

\bibitem{1} See, for example, F. S. Levin, {\it An Introduction to Quantum Theory} (Cambridge University Press, Cambridge, 2002) 
and J. S. Townsend, {\it A Modern Approach to Quantum Mechanics} (University Science Books, Sausalito CA, 2000).

\bibitem{2} See, for example, H. C. Ohanian, {\it Gravitation and Spacetime} (Norton, New York, 1976), Sec. 7.3. 

\bibitem{3}
S. Obradovic, ``The nature of axioms of physical theory," Eur. J. Phys. {\bf 23}, 269--275 (2002).

\bibitem{4} 
J. D. Jackson, {\it Classical Electrodynamics} (John Wiley \& Sons, New York, 1999), 3rd ed., Sec. 6.5.

\bibitem{5}The evaluation at the retarded time must be interpreted unambiguously. The quantity $[\gradv'\cdot \v J]$ means that we first calculate the divergence of the current density at the present time $\gradv'\cdot \v J(\v x',t)$ and in the resulting expression $f_1(\v x',t)$
we replace $t$ by the retarded time $t'=t-R/c$, that is, $f(\v x',t')$. 
Analogously, the quantity $\left[\partial \rho/\partial t\right]$ means that we first calculate the time derivative of the charge density at the present time $\partial \rho(\v x',t)/\partial t$ and in the resulting expression $f_2(\v x',t)$ we replace $t$ by the retarded time $t'=t-R/c$, that is, $f_2(\v x',t')$.
We also note that $\left[\partial \rho/\partial t\right]=\left[\partial \rho/\partial t'\right]$, which follows from the result $\partial \rho(\v x',t')/\partial t=(\partial \rho(\v x',t')/\partial t')(\partial t'/\partial t)=\partial \rho(\v x',t')/\partial t'$.

\bibitem{6} 
This nice interpretation of Eq.~(5) was suggested by an anonymous referee.

\bibitem{7} 
We could also consider sources that vanish sufficiently at infinity, that is, sources of order $O|\v x|^{-2-\delta}$ where $\delta>0$ as $|\v x|$ goes to infinity, $\v x$ being the field point.

\bibitem{8}
A. M. Davis, ``A generalized Helmholtz theorem for time-varying vector fields," Am. J. Phys. {\bf 74}, 72--76 (2006).

\bibitem{9}
J. A. Heras ``Comment on `A generalized Helmholtz theorem for time-varying vector fields,' by A. M. Davis [Am. J. Phys. {\bf 74}, 72--76 (2006)]," Am. J. Phys. {\bf 74}, 743--745 (2006).

\bibitem{10} 
See, for example, D. J. Griffiths, {\it Introduction to Electrodynamics} (Prentice Hall, Englewood, NJ, 1999), 3rd ed., p. 346.

\bibitem{11}
A vector that is equivalent to Eq.~(A17) can be found in R. B. McQuistan, {\it Scalar and Vector Fields: A Physical
Interpretation} (John Wiley \& Sons, New York, 1965), see Eq.~(12.28), which contains the superfluous 
term $(-4\pi/c)R\delta(\v x-\v x')\v\partial [\v J]/\partial t$ that vanishes for $\v x\not=\v x'$ because of the delta function and also for $\v x=\v x'$ because this equality implies $R=0$.

\bibitem{12}
M. Le Bellac and J. M. Levy-Leblond, ``Galilean electromagnetism," Nuovo Cimento B{\bf 14}, 217--233 (1973); M. Jammer and J. Stachel, ``If Maxwell had worked between Amp\'ere and Faraday: An historical fable with a pedagogical moral," Am. J. Phys. {\bf 48}, 5--7 (1980); J. A. Heras, ``Instantaneous fields in classical electrodynamics," Europhys. Lett. {\bf 69}, 1--7 (2005).

\end{thebibliography}
\end{document}